\documentclass[aps,prl,twocolumn,superscriptaddress]{revtex4}

\usepackage{amsmath,amssymb}
\usepackage{graphicx}
\usepackage{times}

\begin{document}
\title{Spin liquid close to a quantum critical point in Na$_4$Ir$_3$O$_8$}
\author{Yogesh Singh}
\affiliation{Indian Institute of Science Education and Research (IISER) Mohali, Knowledge city, Sector 81, Mohali 140306, India}
\affiliation{I. Physikalisches Institut, Georg-August-Universit\"at G\"ottingen, D-37077, G\"ottingen, Germany}
\author{Y. Tokiwa}
\affiliation{I. Physikalisches Institut, Georg-August-Universit\"at G\"ottingen, D-37077, G\"ottingen, Germany}
\author{J. Dong}
\affiliation{I. Physikalisches Institut, Georg-August-Universit\"at G\"ottingen, D-37077, G\"ottingen, Germany}
\author{P. Gegenwart}
\affiliation{I. Physikalisches Institut, Georg-August-Universit\"at G\"ottingen, D-37077, G\"ottingen, Germany}
\date{\today}

\begin{abstract}
Na$_4$Ir$_3$O$_8$ is a candidate material for a 3-dimensional quantum spin-liquid on the hyperkagome lattice. We present thermodynamic measurements of heat capacity $C$ and thermal conductivity $\kappa$ on high quality polycrystalline samples of Na$_4$Ir$_3$O$_8$ down to $T = 500$~mK and $75$~mK, respectively.  Absence of long-range magnetic order down to $T = 75$~mK strongly supports claims of a spin-liquid ground state.  The constant magnetic susceptibility $\chi$ below $T \approx 25$~K and the presence of a small but finite linear-$T$ term in $C(T)$ suggest the presence of gapless spin excitations.  Additionally, the magnetic Gr$\ddot{\rm{u}}$neisen ratio shows a divergence as $T \rightarrow 0$~K and a scaling behavior which clearly demonstrates that Na$_4$Ir$_3$O$_8$ is situated close to a zero-field QCP.

\end{abstract}

\maketitle

In geometrically frustrated materials long range order is prohibited and a liquid-like ground state of spins, the spin-liquid (SL) state, can occur.
Since Anderson's proposal of a valence bond state for a triangular lattice \cite{Anderson1973}, search for such quantum spin liquids (QSL) has been pursued intensely.  In the past decade several new materials have been proposed as SL candidates (see recent reviews \onlinecite{Mila2000, Ramirez2008, Balents2010}).  A small spin and quasi-low-dimensionality are considered ingredients which can lead to a SL state.  Indeed, most of the proposed QSL candidates are materials with $S = 1/2$ moments sitting on quasi-low-dimensional structures.  These include the 2-dimensional (2D) triangular lattice organic compounds $\kappa$-(BEDT-TTF)$_2$Cu$_2$(CN)$_3$ \cite{Yamashita2008, Yamashita2009} and EtMe$_3$Sb[Pd(dmit)$_2$]$_2$ \cite{Yamashita2010}, the 2D kagome lattice inorganic materials ZnCu$_3$(OH)$_6$Cl$_2$, \cite{Helton2007, Han2012} and BaCu$_3$V$_2$O$_8$(OH)$_2$ \cite{Okamoto2009}. 
The QSL states in these and other materials may possess exotic elementary excitations which obey either fermionic or bosonic statistics and have gapped or gapless energy dispersion \cite{Balents2010}.

Whether a QSL can exist in 3D, where quantum effects are reduced and there are more ways in which frustration can be relieved, is still an open question \cite{Balents2010}.  In this context, recently a new 3D material Na$_4$Ir$_3$O$_8$ with effective spins $S = {1\over 2}$ on a frustrated hyperkagome lattice has been proposed as a candidate QSL \cite{Okamoto2007}.  Using thermodynamic measurements at $T \geq 2$~K it was shown that Na$_4$Ir$_3$O$_8$ does not order inspite of strong antiferromagnetic interactions ($\theta = - 650$~K).  The magnetic specific heat showed a bump around 30~K, a temperature much smaller than $\theta$, and
displays a low temperature power-law dependence with exponent close to $2$ \cite{Okamoto2007} similar to several of the 2D materials listed above \cite{Helton2007,Okamoto2009}. 
Subsequently using classical and semi-classical spin-models of Heisenberg spins on a hyperkagome lattice the ground state was found to be highly degenerate and either a classical spin nematic order with long range dipolar spin correlations is chosen at $T \approx 1$~K by an order-by-disorder mechanism \cite{Hopkinson2007} or a 120$^o$ coplanar magnetically ordered state \cite{Lawler2008a} were predicted.  This coplanar magnetic order was found to give way, through a quntum phase transition, to a gapped topological $Z_2$ 'bosonic' spin liquid phase (characterized by the absence of long-range dipolar order) when quantum fluctuations are turned on \cite{Lawler2008a}.  The elementary excitations of this gapped spin liquid state were predicted to be chargeless $S~=~1/2$ spinons \cite{Lawler2008a}.  The prediction of a gapped spin liquid is however at odds with experiments which gave a Sommerfeld's coefficient $\gamma \approx 2$~mJ/Ir~mol~K$^2$ suggesting gapless spin excitations unless the gap was vanishingly small or had nodes.  In a completely different line of approach a Fermionic spin liquid model was developed \cite{Zhou2008, Lawler2008b}.  This led naturally to gapless spin liquids as stable phases.  These spin liquids had a Fermi surface of chargeless spinons which at low temperatures was found to be unstable to the formation of paired states with line nodes \cite{Zhou2008, Lawler2008b}.  This was suggested as a natural explanation of the power law heat capacity at low-$T$ \cite{Zhou2008, Lawler2008b}.   Inclusion of spin-orbit coupling was shown to relieve the frustration in some regions of parameter space and might lead to ordering at low temperatures \cite{Chen2008}.  

Thus there are predictions of both magnetically ordered or (Bosonic or Fermionic) spin-liquid ground states.  As far as low energy excitations are concerned, proposals for both a gapped or gapless state have been made.  Proximity to a Quantum critical point (QCP) has also been suggested \cite{Podolsky2009}. The true ground state of Na$_4$Ir$_3$O$_8$ is thus, still an open question. 

To clarify the true ground state (ordered or spin-liquid), to address the questions about the nature of the low-energy excitations (gapped or gapless), and to look for proximity to a QCP we have performed magnetic susceptibility ($\chi$), heat capacity (C), thermal conductivity ($\kappa$), and magnetocaloric effect [MCE = $(dT/dH)_S)$] measurements down to very low temperatures.  Our main results are: (i) neither $C(T)$ ($T\geq 500$~mK) nor $\kappa(T)$ ($T\geq 75$~mK) show signature of long range magnetic order providing fresh evidence for a spin-liquid,  (ii) both $\chi$ and $C/T$ tend to finite values as $T \rightarrow 0$~K suggesting presence of gapless excitations although $\kappa/T \rightarrow 0$ as $T \rightarrow 0$~K which we argue is most likely due to the polycrystalline nature of the sample,  and (iii) the magnetic Gr$\ddot{\rm{u}}$neisen parameter [$\Gamma_H = {1\over T} ({dT \over dH})_S$] shows a divergence and the data at various applied magnetic fields show a scaling behavior reminiscent of heavy-Fermion materials situated close to a quantum critical point (QCP) \cite{Yoshi2013}.  All the results above demonstrate that \emph{ Na$_4$Ir$_3$O$_8$ is a gapless quantum spin-liquid proximate to a zero-field QCP}.

Polycrystalline samples of Na$_4$Ir$_3$O$_8$ and Na$_4$Sn$_3$O$_8$ were synthesized using standard solid state reaction at $T = 1000^o$C\@.
Measurements down to $T = 2$~K were done on a Quantum Design PPMS.  The low temperature $C$ and $\Gamma_H$ were measured in a dilution refrigerator (DR) \cite{Tokiwa2011}. Low temperature $\kappa$ was measured in a DR, using a standard four-wire steady-state method with two RuO$_2$ chip thermometers, calibrated in situ against a reference RuO$_2$ thermometer.  Electrical transport (not shown) on sintered pellets showed insulating behavior with an activation gap of $\approx 500$~K\@. 

\begin{figure}[t]
\includegraphics[width=3 in]{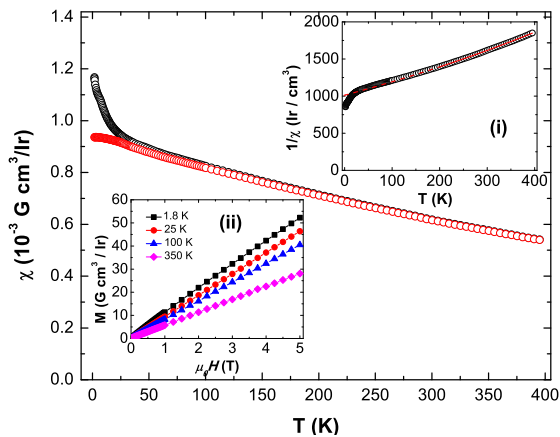}
\caption{(Color online) $\chi(T)$ for Na$_4$Ir$_3$O$_8$ between $T = 1.8$~K and 400~K\@ before (black/open circles) and after (red/inverted triangles) subtraction of an impurity term (see text for details).  Inset~(i) shows the $1/\chi(T)$ data.  The curve through the data is a fit by $\chi = \chi_0 + C/(T-\theta)$.  The inset~(ii) shows $M(H)$ at various $T$. 
\label{Fig-chi}}
\end{figure}

Figure~\ref{Fig-chi} shows $\chi$ (open circles) versus $T$ between $1.8$~K and $400$~K\@.  The data below $T \approx 25$~K show a sharp upturn suggesting contribution from a small amount of paramagnetic impurities.  A fit to this low-$T$ upturn by a Curie-Weiss behavior $\chi = {C \over T-\theta}$ shows that the upturn is well accounted for by $\approx$ 1\% $S = 1/2$ impurities.  Subtracting this from the total $\chi$ we obtain the red curve (inverted triangles) in Fig.~\ref{Fig-chi}.  The corrected $\chi$ saturates to a large and $T$ independent value $\chi \approx 9.4 \times 10^{-4}$~G~cm$^3$/Ir~mol below $T = 25$~K, as it would in a metal, suggesting gapless excitations in Na$_4$Ir$_3$O$_8$.  Figure~\ref{Fig-chi}~inset~(i) shows the $1/\chi(T)$ data.  The data between $T = 300$~K and $400$~K were fit by the Curie-Weiss expression $\chi = \chi_0 + {C\over T-\theta}$.  The fit shown in inset~(i) of Figure~~\ref{Fig-chi}(a) as the curve through the data gave the values $\chi_o = 6.8(5) \times 10^{-5}$~cm$^3$/mol/Ir, $C = 0.43(2)$~cm$^3$~K/mol/Ir, and $\theta = - 512(6)$~K\@.   Assuming a g-factor $g = 2$ the above value of $C$ gives an effective magnetic moment $\mu_{eff} = 1.85(5)~\mu_{\rm B}$ which is close to that expected for $S_{eff} = 1/2$.  The inset~(ii) in Figure~\ref{Fig-chi} shows the magnetization $M$ versus magnetic field $H$ data at various $T$.  The $M(H)$ is linear at all $T$ indicating the absence of any ferromagnetic or large amount of paramagnetic impurity in the sample consistent with the less than 1\% impurities estimated above.  

\begin{figure}[t]
\includegraphics[width=3in]{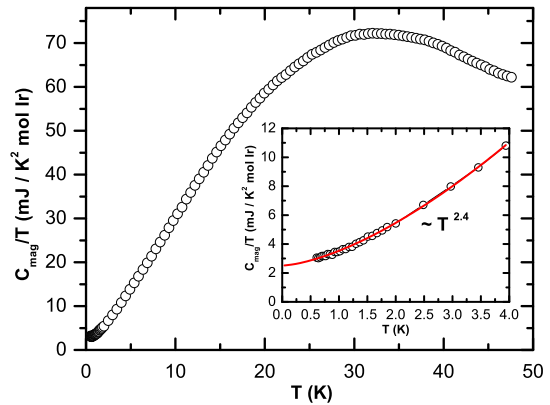}
\caption{(Color online) $C_{mag}/T$ versus $T$ between $T = 0.5$~K and 50~K\@,  (inset) data below $T = 4$~K\@.  The solid curve through the data is a fit by $C = \gamma T + \beta T^n$ (See text for details).  
\label{Fig-CP-LT}}
\end{figure}

The above results of $\chi(T)$ indicates the presence of large antiferromagnetic interactions and the absence of long-range magnetic order down to $T = 1.8$~K\@.  In addition, the $\chi$ is shown to saturate to a $T$-independent value at low temperature.  These results are consistent with what was observed earlier \cite{Okamoto2007} although our Weiss temperature $\theta \approx -500$~K is significantly smaller than the value -650~K estimated in that report.  We have used higher temperature ( $T \leq 400$~K) $\chi(T)$ data to estimate $\theta$ which is probably more reliable given the large value of $\theta$.  

In Fig.~\ref{Fig-CP-LT} we show the magnetic heat capacity divided by temperature $C_{mag}/T$ versus $T$ for Na$_4$Ir$_3$O$_8$ between $T = 0.5$~K and 50~K\@.  $C_{mag}$ was obtained by subtracting the $C(T)$ data of Na$_4$Sn$_3$O$_8$ from the data of Na$_4$Ir$_3$O$_8$.  $C_{mag}/T$ shows a broad anomaly peaked at $T \approx 30$~K\@ consistent with what was observed earlier \cite{Okamoto2007}.  We did not find any signature of long range magnetic order down to $T = 0.5$~K in our heat capacity data.  An entropy of $Rln2$ expected for $S = 1/2$ moments is recovered at $T = 50$~K, a temperatures a little above the anomaly in $C_{mag}$.  This suggests that the broad anomaly in $C_{mag}$ could be the onset of short ranged magnetic order.  However, it could also be an indication of something more exotic like a crossover to a correlated spin-liquid like state from a thermally disordered high temperature state dominated by nearest-neighbor interactions.  A similar anomaly has been observed at $T = 6$~K in the organic spin-liquid candidate $\kappa$-(BEDT-TTF)$_2$Cu$_2$(CN)$_3$ \cite{Yamashita2008}.  In the case of the latter compound, NMR and $\mu$SR measurements below this temperature did not detect signature of internal fields indicating magnetic order was not causing the anomaly.  

$C_{mag}$ below $T = 4$~K is shown in the inset of Fig.~\ref{Fig-CP-LT}.  These data were fit by the expression $C = \gamma T + \beta T^n$.  We were able to obtain an excellent fit, shown as the solid curve through the data in the inset of Fig.~\ref{Fig-CP-LT}, with the values $\gamma = 2.5(2)$~mJ/K$^2$~mol~Ir, $\beta = 3.0(1)$~mJ/mol~K$^4$, and $n = 2.41(2)$.  An exponent $n$ between $2$ and $3$ was also observed earlier for data measured down to $T = 2$~K \cite{Okamoto2007}.  One of the mechanisms leading to this power-law behavior in $C$ could be the predicted line-nodes in a spinon Fermi surface \cite{Zhou2008, Lawler2008b}.  The most important conclusion from our low temperature $C$ data is the presence of a small but finite linear-$T$ term.  In this insulating material where charge excitations are gapped, the linear term is expected to arise due to gapless spin excitations.  We note that we did not observe signature of any low-lying Schottky anomaly associated with free paramagnetic impurities.  

\begin{figure}[t]
\includegraphics[width=3 in]{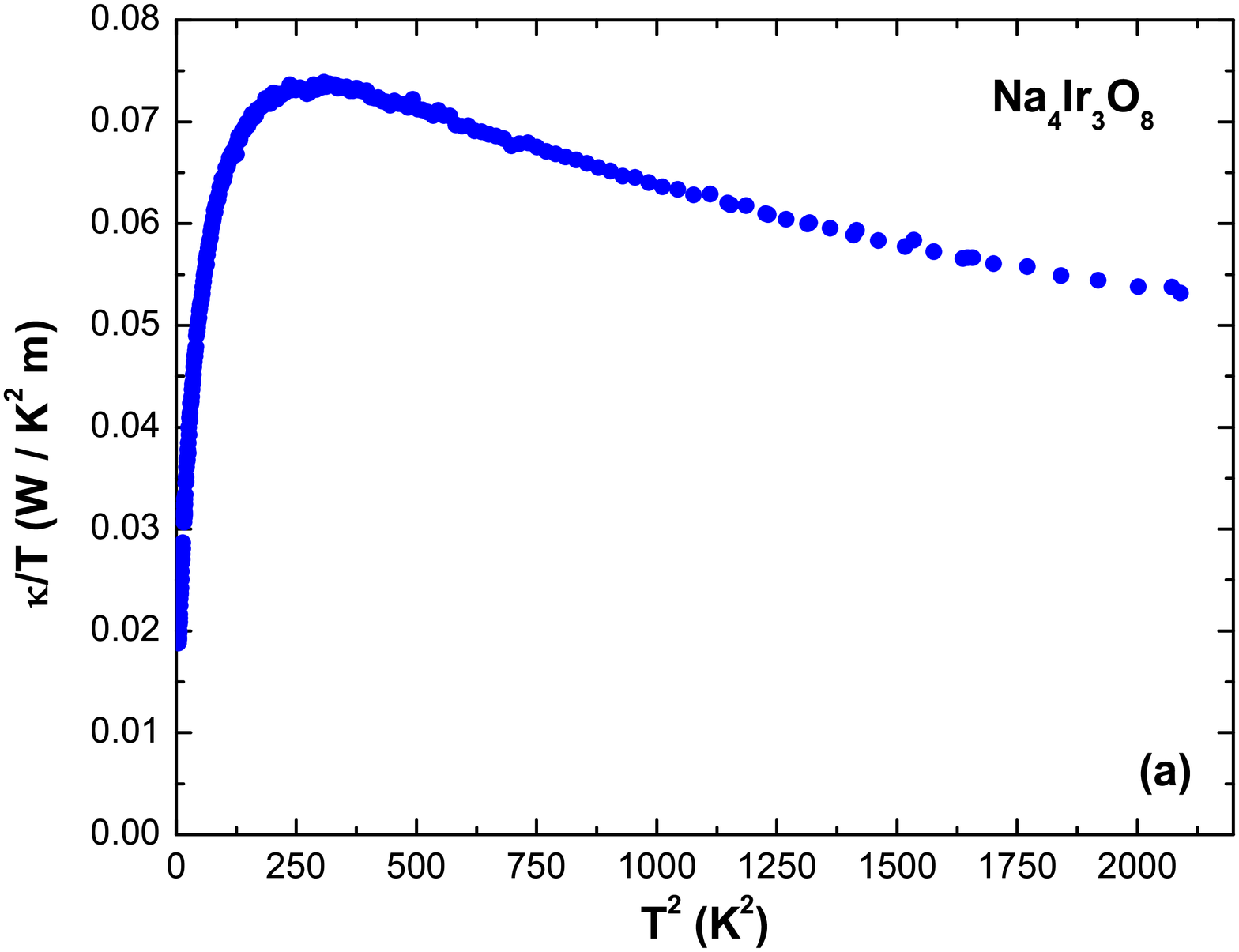}
\includegraphics[width=3 in]{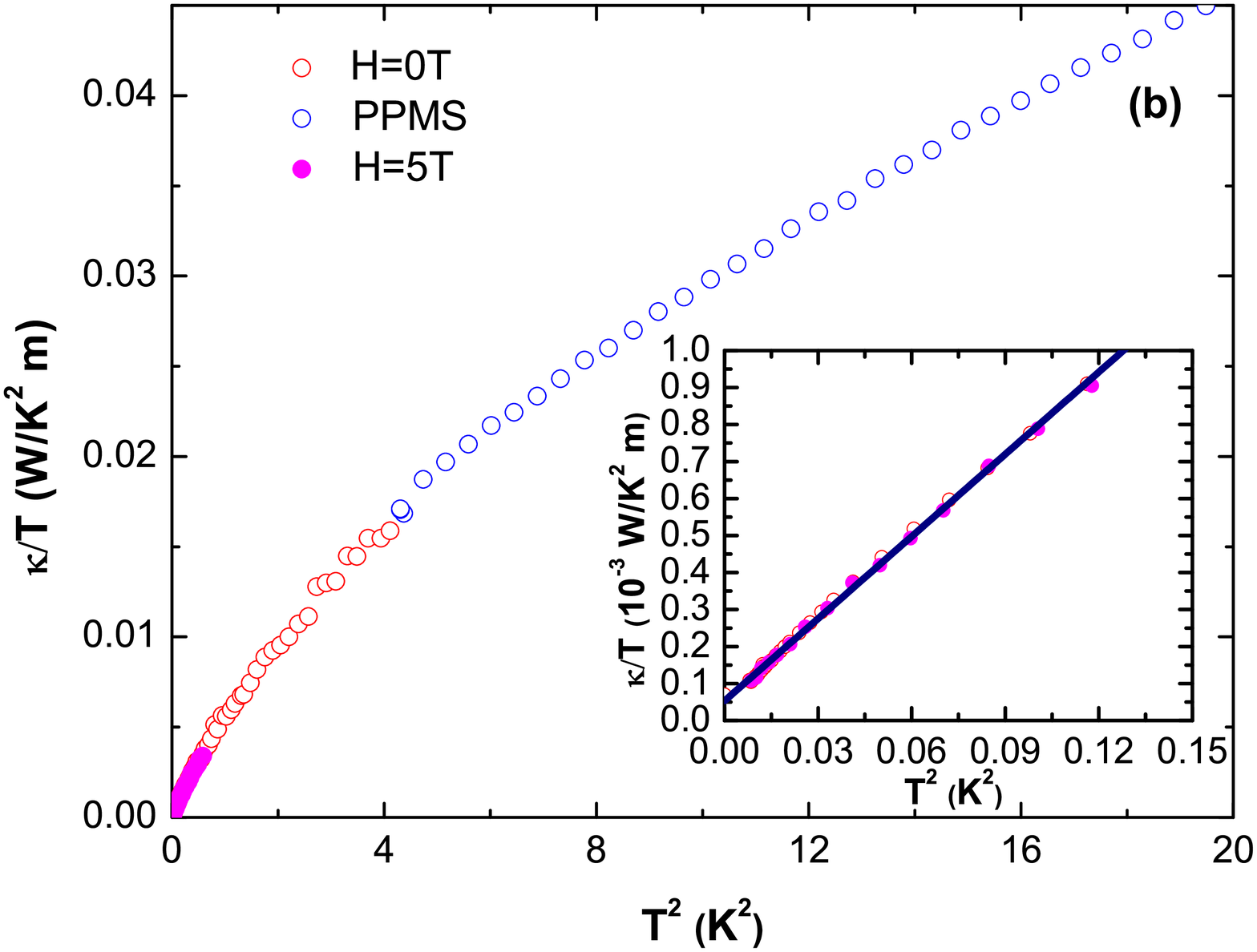}
\caption{(Color online) (a) $\kappa/T$ versus $T^2$ measured at $H = 0$~T in a PPMS.  (b) $\kappa/T$ versus $T^2$ at $H = 0$~T and $H = 5$~T measured in a dilution fridge.  (inset) $\kappa/T$ versus $T^2$ data for $T \leq 0.4$~K to highlight the absence of a significant linear-$T$ term.  
\label{Fig-kappa}}
\end{figure}

Figure~\ref{Fig-kappa} shows the thermal conductivity ($\kappa$) data for Na$_4$Ir$_3$O$_8$.  Figure~\ref{Fig-kappa}(a) shows the zero field $\kappa/T$ versus $T^2$ data between $T = 2$~K and $47$~K\@.  The temperature dependence of $\kappa$ is typical of insulators with a dominant lattice contribution.  The dilution fridge data between $T = 75$~mK and 2~K is plotted along with the PPMS data in Figure~\ref{Fig-kappa}(b).  Both sets of data match quite well.  A signature of long range magnetic ordering (spin-glass) would be an abrupt (gradual) increase in $\kappa$ due to freezing out of the spin disorder scatterring below the ordering (freezing) temperature.  The $\kappa(T)$ data in Figure~\ref{Fig-kappa} do not show any signature of magnetic ordering or spin freezing  providing fresh evidence to support claims of a spin-liquid state in Na$_4$Ir$_3$O$_8$.

The data below $T \approx 0.4$~K is shown in Figure~\ref{Fig-kappa}(b)~inset.  The data extrapolates to $\kappa(0)/T = 6.3(9)\times10^{-2}$~mW~K$^2$/m at $T = 0$~K\@.  This is independent of magnetic field as can be seen from the $H = 5$~T data plotted along with the $H = 0$~T data in Figure~\ref{Fig-kappa}(b).  
The extrapolated $T = 0$~K value of $\kappa(0)/T$ is tiny and within errors is indistinguishable from zero.  The fact that our samples are polycrystalline leads to two issues.   At low temperatures when the phonon mean free path becomes large, scattering from grain boundaries becomes important and phonon contribution to $\kappa$ deviates from the diffuse scattering limit $T^3$ behavior.  Thus, there is no clear way of estimating and subtracting the phonon contribution to $\kappa$ and getting out a magnetic or itinerant contribution.  For the same reason $C$ and $\kappa$, which would otherwise be related, can not be compared in this case.  In our samples the measured thermal conductivity, specially at low temperatures, will be lowered from its intrinsic value due to thermal resistance coming from grain boundary scattering.  So $\kappa(0)/T = 6.3(9)\times10^{-2}$~mW~K$^2$/m can be considered a lower limit.

\begin{figure}[t]
\includegraphics[width=3 in]{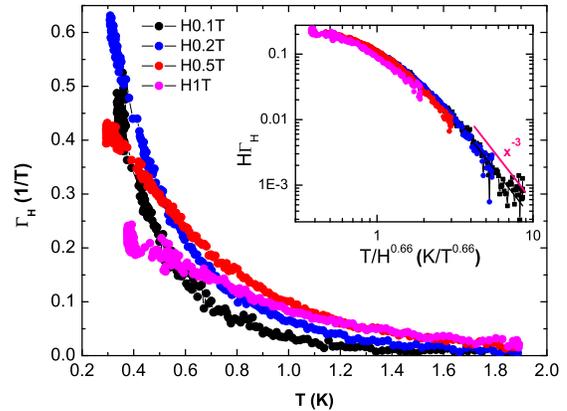}
\caption{(Color online) Magnetic Gr$\ddot{\rm{u}}$neisen parameter $\Gamma_H$ versus $T$, (inset) scaling plot, $H \Gamma_H$ vs ${T/H^{\nu z}}$ with $\nu z= 2/3$.  At large $x = T/H^{\nu z}$ we observe a $x^{-3}$ divergence.  
\label{Fig-MCE}}
\end{figure}

We next present evidence for the proximity of Na$_4$Ir$_3$O$_8$ to a zero-field quantum critical point (QCP).  Figure~\ref{Fig-MCE} shows the magnetic Gr$\ddot{\rm{u}}$neisen ratio $\Gamma_H$ versus $T$ data between $T = 0.3$~K and $T = 2$~K at various magnetic fields.  A clear divergence of the $\Gamma_H$ is observed at low fields.  The divergence of the Gr$\ddot{\rm{u}}$neisen parameter is a typical signature of proximity to a field tuned QCP \cite{Gegenwart2010}.  Similar divergences have been observed at the field-induced QCP in the heavy fermion material YbRh$_2$Si$_2$ \cite{Tokiwa2009}.  The divergence of $\Gamma_H$ is suppressed at higher fields indicating that at these fields the system is being tuned away from the QCP.  

Assuming only a single diverging time scale governs the critical phenomena, the critical free energy, $F_{cr}$ with magnetic field as control parameter can be expressed by $F_{cr}=aT^{(d+z)/z}\phi(bh/T^{1/\nu z})$, where $h = H-H_c$~\cite{Zhu2003, Garst2005,Lohneysen2007} and $a$ and $b$ are non-universal constants.  Thus, thermodynamic quantities, that are derivative(s) of free energy, are expected to collapse as a function of $T/h^{\nu z}$.  Here, we present in Fig.~4 inset an excellent collapse of our data, when they are rescaled in a form $H\Gamma_H$ versus $T/H^{0.67}$, indicating $H_c = 0$ and implying $\nu z = 2/3$.  These data demonstrate that Na$_4$Ir$_3$O$_8$ is situated at or close to a zero-field QCP.

By expanding the dimensionless argument $y=(bh)/T^{1/\nu z}$ in $\phi$ for high temperatures, $y\ll 1$, we obtain $\phi(y)\approx \phi(0)+1/2\phi''(0)y^2$, where
the $y$-linear term ($H$-linear term) vanishes, because it would correspond to a spontaneous magnetization.  Now, $\Gamma_{\rm H}=[\partial^2F_{cr}/(\partial T\partial H)]/[T\partial^2F_{cr}/\partial T^2]$, yielding $\Gamma_{\rm H}H\sim [b'T^{2/\nu z}/H^2+1]^{-1}$. Here $b'$ is another non-universal constant.  From this we expect $\Gamma_{\rm H}H\sim (T/H^{\nu z})^{-2/\nu z}$ at high $T$.  The high $T$ exponent, $-3=-2/\nu z$ (solid line in Fig.~4~inset), is in perfect agreement with $\nu z$ = 0.67, obtained from the argument of scaling function, $\phi_T(T/H^{\nu z})$.
  
We note that $\Gamma_H$ for $H ~=~ 0.2$~T is larger than that for $H ~=~0.1$~T in Fig.~\ref{Fig-MCE}, which is counter-intuitive to expectation for a zero-field QCP.  A crossing between two curves of $\Gamma_H$ at different $H$ is expected as a function of $T$ for $H>H_c$~\cite{Garst2005}.  This is due to a cross-over between high-temperature $H$-linear [since $\Gamma_HH\sim (T/H^{2/3})^{-3}$] and low-temperature 1/$H$ regimes.  For example, the curve for $H = 0.1$~T crosses with that for $H = 1$~T at 0.55~K and the curve for $H = 0.5$~T at $T = 0.4$~K\@.  The crossing temperature decreases with decreasing field. From the scaling plot, the $\Gamma_H$ curves for $H = 0.1$ and 0.2~T are expected to cross below 0.3~K, which is the lowest temperature in our experiment.

We have shown that Na$_4$Ir$_3$O$_8$ is a fascinating material with several novel properties.  We have presented three main results here. (i) The absence of long range magnetic (or other) order down to $T = 75$~mK strongly supporting a spin-liquid ground state.  (ii) Suggestion of gapless excuitations from finite terms in $\chi$ and $C$ in the limit $T\rightarrow 0$ :  Even though Na$_4$Ir$_3$O$_8$ is shown to be a Mott insulator from transport and magnetism, the $\chi$ becomes $T$-independent below $T \approx 25$~K and the  heat capacity shows a small but finite linear-$T$ term at the lowest temperatures.  Since the charge excitations are gapped, the above implies the \emph{presence of gapless spin excitations in Na$_4$Ir$_3$O$_8$}.  The thermal conductivity however, extrapolates to a tiny value $\kappa/T \approx 6.5 \times 10^{-2}$~mW/K$^2$~m.  The values of the linear-$T$ terms in $C$ and $\kappa$ are both small.  
If we compare with the two-dimensional organic spin-liquid candidates ET-dmit ($\kappa/T \approx 180$~mW/K$^2$~m, $\gamma \approx 20$~mJ/mol~K$^2$)  and kappa-(BEDT-TTF)$_2$Cu$_2$(CN)$_3$ ($\kappa / T \rightarrow 0$ \cite{Yamashita2009}, $\gamma \approx 20$~mJ/mol~K$^2$ \cite{Yamashita2008}) we find that the case of Na$_4$Ir$_3$O$_8$ seems similar to the latter material where a relatively large $\gamma$ is found but is not consistent with the vanishing linear-$T$ term in thermal conductivity, both of which were obtained from measurements on single crystals.  

Recent zero-field $\mu$SR measurements on kappa-(BEDT-TTF)$_2$Cu$_2$(CN)$_3$ reveal a microscopic phase separation below about $3$~K into gapless paramagnetic nano-islands embedded in a sea of a gapped singlet phase \cite{Nakajima2012}.  This was proposed as a resolution to the conflicting results of $C$ and $\kappa$ because the heat capacity will measure both contributions (giving a net gapless result) whereas the thermal conduction between gapless islands will be cutoff by the gapped sea around the islands \cite{Nakajima2012}.  The absence of a significant $\kappa(0)/T$ for Na$_4$Ir$_3$O$_8$ could thus be a result of something exotic.  Future microscopic measurements like NMR or $\mu$SR would be useful in this regard.  We would nevertheless like to stick our necks out and state that the evidence from $C$ and $\chi$ at low $T$ strongly suggest gapless excitations.  These could arise e.g. from a Fermi surface of spinons and in analogy with metals the small values of $\gamma$ and $\kappa(0)/T$ might therefore suggest a small spinon density of states.  This would be consistent with theoretical predictions of three small spinon Fermi surface pockets for Na$_4$Ir$_3$O$_8$ \cite{Zhou2008}.  These are however still speculations at this stage and direct evidence of the existence of a spinon FS in Na$_4$Ir$_3$O$_8$ is hardly established.

(iii) The third main result of our work is proximity to a QCP: The divergence of $\Gamma_H$ points to quantum criticality.  The quantum criticality must be associated with an instability of the system to some nearby (possibly magnetic) ordered state.  Note, that $\Gamma_H$ of fluctuating paramagnetic moments is independent of $T$.  Therefore, the observed scaling cannot be explained by an impurity contribution and provides strong evidence for quantum critical behavior in Na$_4$Ir$_3$O$_8$.

\paragraph{Acknowledgments.--} YS acknowledges support from DST, India and acknowledges University of Goettingen, for a guest professorship during the summer of 2012.  Work at Goettingen supported by the DFG reserach group 960 (Quantum Phase Transitions), the Helmholtz Virtual Institute 521 ("New states of matter and their excitations") and the Alexander-von-Humboldt foundation.  This work was supported in part by the National Science Foundation under Grant No. PHYS-1066293 and the hospitality of the Aspen Center for Physics.

\paragraph{Note added:--} A recent muSR measurement \cite{kwang-yong} on powder Na$_4$Ir$_3$O$_8$ samples prepared by YS using methods similar to the ones used for samples in our manuscript have shown the absence of long-range magnetic ordering or spin freezing at 5 K strongly supporting a SL ground state at least down to these temperatures.

\end{document}